\documentclass[reprint,twocolumn,superscriptaddress,secnumarabic,amssymb, nobibnotes, aps, prb]{revtex4-1}

\usepackage{amsmath}    
\usepackage{graphicx}    
\usepackage{verbatim}    
\usepackage{color}           
\usepackage{subfigure}   
\usepackage{hyperref}    
\usepackage{braket}
\usepackage{physics}
\usepackage{xfrac}
\usepackage{xspace}
\usepackage{booktabs}
\usepackage[stable]{footmisc}
\usepackage{makecell}
\newcommand{\amt}{$\alpha$-MnTe\xspace}
\newcommand{\bk}{\mathbf{k}}
\newcommand{\bki}{\mathbf{k}_{\rm in}}
\newcommand{\bko}{\mathbf{k}_{\rm out}}

\newcommand{\bM}{\mathbf{M}}

\newcommand{\mU}{\mathcal{U}}
\newcommand{\mT}{\mathcal{T}}

\newcommand{\bs}{\mathbf{s}}
\newcommand{\bL}{\mathbf{L}}
\newcommand{\bR}{\mathbf{R}}
\newcommand{\cb}{[11\bar{2}0]}
\newcommand{\ca}{[1\bar{1}00]}

\newcommand{\be}{\boldsymbol{\varepsilon}}
\newcommand{\bnu}{\boldsymbol{\nu}}

\def\bL{\mathbf{L}}

\def\hT{\hat{T}}

\begin{document}
\title{Theory of circular dichroism in resonant inelastic x-ray scattering}

\author{M.~Furo}
\affiliation{Department of Physics and Electronics, Graduate School of Engineering,
Osaka Metropolitan University, 1-1 Gakuen-cho, Nakaku, Sakai, Osaka 599-8531, Japan}
\author{A.~Hariki}
\email[]{hariki@omu.ac.jp}
\affiliation{Department of Physics and Electronics, Graduate School of Engineering,
Osaka Metropolitan University, 1-1 Gakuen-cho, Nakaku, Sakai, Osaka 599-8531, Japan}
\author{J.~Kune\v{s}}
\email[]{kunes@physics.muni.cz}
\affiliation{Department of Condensed Matter Physics, Faculty of
  Science, Masaryk University, Kotl\'a\v{r}sk\'a 2, 611 37 Brno,
  Czechia}

\begin{abstract}

We analyze circular dichroism (CD) in resonant inelastic x-ray scattering (RIXS) in magnetic materials.
We define RIXS-CD as the difference between scattering amplitudes for the right- and left-circularly polarized incoming photons and unpolarized (total) outgoing photons. We employ the impurity approximation, in which the interference between scattering
events on different atoms is neglected. We perform the symmetry analysis of several common antiferromagnetic and altermagnetic 
structures and outline the general approach. The analysis is supported by numerical calculations using atomic model
with realistic crystal fields obtained from first principles. We show that RIXS-CD is distinguished from first-order spectroscopies 
such as x-ray magnetic circular dichroism by insensitivity to the time-reversal symmetry breaking. As a result we find that
RIXS-CD is present in the normal (disordered) state of materials with lower symmetry. 
In antiferromagnets the RIXS-CD is 
invariant under N\'eel vector reversal. In altermagnets and ferromagnets the RIXS-CD spectra for time-reversed states
are, in general, independent except for the special case when there is a unitary symmetry of the Hamiltonian connecting the
two states.

\end{abstract}

\maketitle

\section{Introduction}
X-ray magnetic circular dichroism (XMCD) has recently demonstrated its utility as a tool for characterizing some classes compensated magnets~\cite{Yamasaki20,Kimata21,Sakamoto21,Sasabe21,Sasabe23,Hariki24,Hariki25,Hariki24_ruo2,Hariki24_mnf2}. It enabled mapping out of antiferromagnetic domains~\cite{Amin24}.
Resonant inelastic x-ray scattering (RIXS) a related technique~\cite{Ament11,Mitrano24,deGroot24,groot_kotani}. While experimentally more demanding, RIXS with more control parameters provides a more detailed insight into the excitations of the studied material, e.g., momentum resolution, and is less restricted by selection rules.
RIXS experiments with circularly polarized x-rays have recently been reported for several materials from the currently much studied altermagnetic group~\cite{Takegami25_pre,Jost25_pre,Biniskos25_pre}. The resonant nature of RIXS process makes the interpretation of the measured spectra rather non-trivial requiring some theoretical input~\cite{Fukui01,Kotani05}.
The key question in circular dichroism (CD) experiments concerns the selection rules. When does the effect vanish due to symmetry constraints
and conversely what a finite CD implies about the symmetry of the studied material? The selection rules for XMCD in the dipole approximation are quite restrictive, in particular, it is forbidden in systems with time reversal symmetry.

In this paper we develop the theory of RIXS-CD and present numerical simulations for selected materials. We find that RIXS-CD is much more prolific than XMCD and can be found even in non-magnetic systems with lower symmetry. We employ two approximations. The dipole approximation, which provides the leading order in x-ray absorption, means that the wave vectors of incoming and outgoing photons enter only by selecting the corresponding polarization planes.
The impurity approximation, means that we neglect the interference between the scattering events on different atoms. In the impurity approximation,
the momentum transfer to the electronic excitations, and material final state in general, is neglected and thus dispersion of excitations cannot be described. This may be rather crude for dispersive low-energy excitations such as magnons, which are described as local spin flips. Importantly, 
the final state symmetry in the impurity approximation is higher that in general and thus the derived selection rules are more restrictive. This means that
RIXS-CD may be weakly allowed even when it is forbidden in the present theory.

We start by discussing the time-reversal symmetry. We show that on-resonance RIXS process itself breaks the time-reversal symmetry and thus experiments
performed on time-reversed states are not related to one another. The RIXS spectra are therefore sensitive only to unitary (geometric) symmetries of the system.
Next, we show that the dependence of RIXS-CD on the incoming and outgoing wave vectors in the impurity approximation can be written 
as a sum of an isotropic part depending only on the incoming wave vector and characterized by a pseudovector, analogy of a Hall vector, and an anisotropic part described by a rank-3 tensor. The form of this tensor is determined by the unitary subgroup of the magnetic point group. We derive these for some materials of current interest. 

\section{Theory}

The RIXS intensity for  an initial state $|i\rangle$~\footnote{The initial state projector $\sum_i|i\rangle\langle i|$ may contain
Boltzmann factors at a finite temperature, which are not shown for sake of readebility.} 
is given by Kramers-Heisenberg (KH) formula~\cite{Kramers25,groot_kotani,Ament11}
\begin{equation}
\label{eq:F_basic}
\begin{split}
F(\omega_\text{out},\omega_\text{in})=&\sum_{f,i}\left|\sum_m\frac{\expval{f|\hT_\text{out}|m}\expval{m|\hT_\text{in}|i}}
{E_i-E_m+i\Gamma}\right|^2\\
&\times\delta(E_f-E_i).
\end{split}
\end{equation}
Here, $|m\rangle$, and $|f\rangle$ are the intermediate and final states, respectively.
The energies of the initial state $E_i=\tilde{E}_i+\omega_\text{in}$ and final state $E_f=\tilde{E}_f+\omega_\text{out}$
include the energy of the incoming ($\omega_{\rm in}$) and outgoing ($\omega_{\rm out}$) photon. $\Gamma$ represents the inverse lifetime of the core hole in the intermediate state. Importantly, this term does not vanish even in the limit of an infinite core-hole lifetime in which case one has to take the $\Gamma\rightarrow 0^+$ limit yielding a finite imaginary part if a continuous intermediate-state spectrum is resonant with the initial/final states.
This reflects the irreversible nature of the initial-state decay to the intermediate-state continuum. 
Interference between these transitions and transitions via a virtual (non-resonant) intermediate state,
described by the real part of the denominator, makes the RIXS intensity insensitive to the time-reversal symmetry breaking.

To show this let us compare the RIXS intensity between a given state and a state related by an anti-unitary operation. 
\begin{equation}
    \begin{split}
   & F^\mT=\sum_{f,i}\left|\sum_m\frac{\expval{\mT f|\hT_\text{out}|\mT m}\expval{\mT m|\hT_\text{in}|\mT i}}
{E_i-E_m+i\Gamma}\right|^2\delta_{fi}\\
 &=\sum_{f,i}\left|\sum_m\frac{ \overline{\expval{ f|\mT^{-1}\hT_\text{out}\mT| m}}\ \overline{\expval{m|\mT^{-1}\hT_\text{in}\mT| i}}}
{E_i-E_m+i\Gamma}\right|^2\delta_{fi}.\\
    \end{split}
\end{equation}
Thanks to the complex nature of the denominator there is no linear relationship between the RIXS spectrum of the original and time-reversed system~\footnote{Unless all the terms in the numerator have the same phase, which we find unrealistic.}.
This is different for the unitary transformations. 
\begin{equation}
\label{eq:unitary}
    \begin{split}
   & F^\mU=\sum_{f,i}\left|\sum_m\frac{\expval{\mU f|\hT_\text{out}|\mU m}\expval{\mU m|\hT_\text{in}|\mU i}}
{E_i-E_m+i\Gamma}\right|^2\delta_{fi}\\
 &=\sum_{f,i}\left|\sum_m\frac{ \expval{ f|\mU^{-1}\hT_\text{out}\mU| m}\expval{m|\mU^{-1}\hT_\text{in}\mU| i}}
{E_i-E_m+i\Gamma}\right|^2\delta_{fi}.\\
    \end{split}
\end{equation}
In this case, $F^\mU$ and $F$ are related by rotations of the dipole operators, i.e., $F$ transforms as the fourth power of the vector representation. 
This may be used to relate the spectra from different states, e.g., magnetic domains, or to constrain the form of $F$, if $\mU$ is the symmetry of the system\footnote{This means that projectors $\sum_f |f\rangle\langle f|$, $\sum_m |m\rangle\langle m|$ and
$\sum_i |i\rangle\langle i|$ are invariant under the action of $\mU$.}

The above analysis has an important implication for the RIXS-CD and its relationship to CD in x-ray absorption. The XMCD 
vanishes
in systems with time-reversal symmetry and thus it is typically observed in magnetic systems with broken time-reversal symmetry.
In the RIXS process the time-reversal symmetry by itself cannot make CD vanish. Instead RIXS-CD reflects only the unitary symmetries and their changes across the phase transitions. 

The momentum deposited in the system by the RIXS process substantially reduces the number of unitary symmetries as these must preserve the transferred momentum. Therefore, Eq.~\eqref{eq:unitary} typically provides a relationship between signals in different scattering geometries rather than constraining the RIXS spectra for a given momentum transfer, similar to the little group of a k-point in the band structure theory. However, in the impurity approximation, the unitary symmetries span the point group of the system~\footnote{The unitary subgroup of the magnetic point group in magnetically ordered systems.} and thus
pose stricter restrictions on the RIXS-CD spectra. In some cases the dispersions of specific excitations are weak with impurity model providing an accurate approximation. In other cases, the observation of finite RIXS-CD in geometries for which it is forbidden in the impurity approximation could allow to draw conclusions about the properties of propagating excitations, such as associating RIXS-CD with the chirality of magnons.

\hspace{-1cm}

\subsection{Anderson impurity approximation}
In the impurity approximation the total RIXS amplitude is the sum of contributions of individual atoms. 
The wave vectors $\bk_\text{in}$ and $\bk_\text{out}$ enter only through the polarizations $\be$ and $\bnu$ of the 
incoming and outgoing radiation, respectively.

\begin{figure}
    \centering
    \includegraphics[width=0.9\columnwidth]{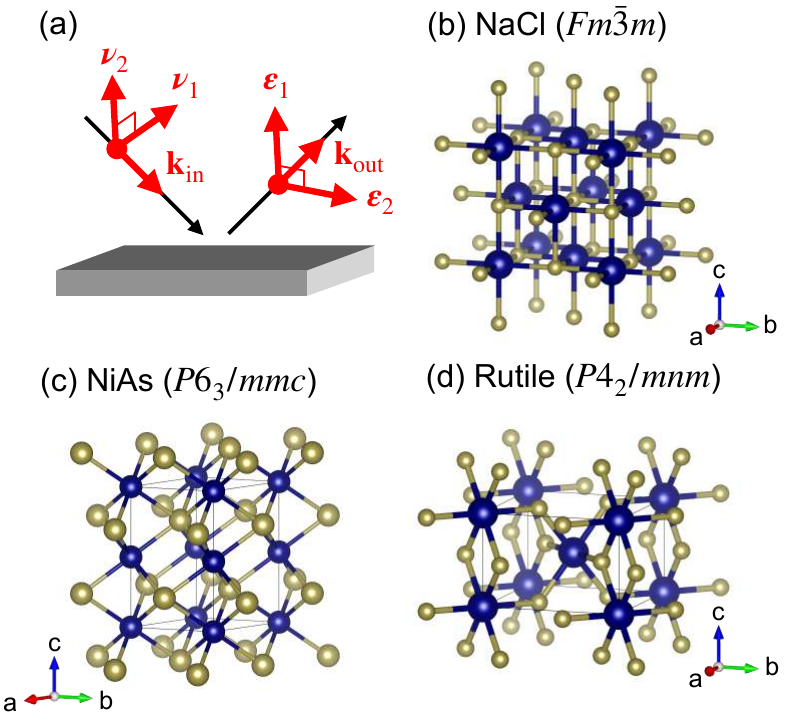}
    \caption{(a) Wave vectors and polarizations of incoming and outgoing x-rays. (b) NaCl structure, (c) NiAs structure, and (d) Rutile structure.}
    \label{fig_geo}
\end{figure}

\begin{widetext}
\begin{equation}
\begin{split}
F(\be,\bnu)=
&\sum_{f,i}\left|\sum_m\frac{\expval{f|\hat{\mathbf{T}}\cdot\be|m}\expval{m|\hat{\mathbf{T}}\cdot\bnu|i}}
{E_i-E_m+i\Gamma}\right|^2
\delta_{fi}\\
&=\left[\sum_{f,i}\sum_{m,m'}
\frac{\expval{i|\hT_\gamma|m'}\expval{m'|\hT_\alpha|f}\expval{f|\hT_\beta|m}\expval{m|\hT_\delta|i}}
{(E_i-E_m+i\Gamma)(E_i-E_{m'}-i\Gamma)}\delta_{fi}\right]
\overline{\varepsilon}_\alpha\varepsilon^{\phantom\dagger}_\beta \overline{\nu}_\gamma\nu^{\phantom\dagger}_\delta
\equiv F_{\alpha\beta,\gamma\delta}\overline{\varepsilon}_\alpha\varepsilon^{\phantom\dagger}_\beta \overline{\nu}_\gamma\nu^{\phantom\dagger}_\delta,
\end{split}
\end{equation}
\end{widetext}
where $\hT_\alpha$ are the cartesian (hermitian) components of the dipole operator on a given atom. Atomic indices are not shown for sake of simplicity.
We consider the geometry of Fig.~\ref{fig_geo} with the circularly polarized x-rays coming along the $\bk_\text{in}$ direction and the total intensity of an X-ray scattered in the $\bk_\text{out}$ direction being recorded. The
 RIXS-CD signal the sum over the two complementary polarization of the outgoing radiation
$\be_1$ and $\be_2$ and the difference of the two incoming circular polarizations $\bnu_1+i\bnu_2$ and $\bnu_1-i\bnu_2$. 
It can be expressed as a multilinear function of $\bk_\text{out}$ and $\bk_\text{in}$
\begin{equation}
\label{eq:tensor}
    \begin{split}
        \Delta F =& F_{\alpha\beta,\gamma\delta}\left(\varepsilon^1_\alpha \varepsilon^1_\beta + \varepsilon^2_\alpha \varepsilon^2_\beta\right)\left((\nu^1_\gamma-i\nu^2_\gamma)(\nu^1_\delta+i\nu^2_\delta)-c.c.
        \right)\\
        =&-2iF_{\alpha\beta,\gamma\delta}\epsilon_{\gamma\delta\eta}
        \left(\delta_{\alpha\beta}-\hat{k}^\text{out}_\alpha\hat{k}^\text{out}_\beta\right) \hat{k}^\text{in}_\eta
        \\
        =& -2i\left(F_{\mu\mu,\gamma\delta}\delta_{\alpha\beta}-F_{\alpha\beta,\gamma\delta}\right)\epsilon_{\gamma\delta\eta}
        \hat{k}^\text{out}_\alpha\hat{k}^\text{out}_\beta\hat{k}^\text{in}_\eta\\
        =& G_{\alpha\beta,\eta}\hat{k}^\text{out}_\alpha\hat{k}^\text{out}_\beta\hat{k}^\text{in}_\eta.
    \end{split}
\end{equation}
Here we have used that $\{\be_1$, $\be_2, \hat{\mathbf{k}}^\text{out}\}$ 
and $\{\bnu_1$, $\bnu_2, \hat{\mathbf{k}}^\text{in}\}$
form orthonormal bases. Therefore $\varepsilon^1_\alpha \varepsilon^1_\beta + \varepsilon^2_\alpha \varepsilon^2_\beta+\hat{k}^\text{out}_\alpha\hat{k}^\text{out}_\beta=\delta_{\alpha\beta}$ and $\hat{k}^\text{in}_\eta = \epsilon_{\eta\gamma\delta} \nu^1_\gamma\nu^2_\delta$.
The RIXS-CD dependence on $\bk_\text{out}$ and $\bk_\text{in}$ is thus described by described by rank-3 tensor, which transforms as a symmetric product of vectors in the first two indices and as a pseudovector in the third index. 

Similarly, we can express the $\bk_\text{out}$- and $\bk_\text{in}$-dependence of the total unpolarized RIXS intensity, in which case
we sum over the complementary polarizations for both incoming and outgoing x-rays.
\begin{equation}
\label{eq:tensor}
    \begin{split}
        F_\text{tot} =& F_{\alpha\beta,\gamma\delta}
        \left(\varepsilon^1_\alpha \varepsilon^1_\beta + \varepsilon^2_\alpha \varepsilon^2_\beta\right)
        \left(\nu^1_\alpha \nu^1_\beta + \nu^2_\alpha \nu^2_\beta\right)\\
        =&F_{\alpha\beta,\gamma\delta}
        \left(\delta_{\alpha\beta}-\hat{k}^\text{out}_\alpha\hat{k}^\text{out}_\beta\right)
        \left(\delta_{\gamma\delta}-\hat{k}^\text{in}_\gamma\hat{k}^\text{in}_\delta\right)\\
        =& \left (F_{\mu\mu,\nu\nu}\delta_{\alpha\beta} \delta_{\gamma\delta}
        - F_{\mu\mu,\gamma\delta}\delta_{\alpha\beta}-F_{\alpha\beta,\nu\nu} \delta_{\gamma\delta}
        + F_{\alpha\beta,\gamma\delta}\right)\times\\
        &\hat{k}^\text{out}_\alpha\hat{k}^\text{out}_\beta
        \hat{k}^\text{in}_\gamma\hat{k}^\text{in}_\delta
        \\
        =&
        G'_{\alpha\beta,\gamma\delta}\hat{k}^\text{out}_\alpha\hat{k}^\text{out}_\beta
        \hat{k}^\text{in}_\gamma\hat{k}^\text{in}_\delta.
    \end{split}
\end{equation}

Since there is no interference between processes on different atomic sites, the unitary symmetries relevant in the impurity approximation form the  unitary subgroup of the magnetic point group. The RIXS-CD tensor $G_{\alpha\beta,\gamma}$ transforms as a symmetric product in the first and second index and as a pseudovector (anti-symmetric product) in the third index. The total RIXS tensor $G'_{\alpha\beta,\gamma\delta}$ transforms as a symmetric product in the first and second index, and in the third and fourth index. It is reminiscent of the elasticity tensor, however, there is no symmetry corresponding to the exchange of $\hat{\mathbf{k}}^\text{in}$ and $\hat{\mathbf{k}}^\text{out}$.

\begin{table*}[ht]
\centering
\caption{RIXS-CD tensor $G_{\alpha\beta,\eta}$ for selected magnetic compounds: $\hat{\mathbf{L}}$ denotes the orientation of the magnetic moments, $M$ is the magnetic point group and $H$ is the unitary subgroup (halving subgroup). The Cartesian coordinates for the $G_{\alpha\beta,\eta}$ tensor coincide with the lattice vectors if orthogonal. In the other cases $z\parallel c$ and the orientation of the $x$ axis is indicated in the last column. 
}
\label{tab:symmetry-data}
\renewcommand{\arraystretch}{1.1}
\begin{tabular}{l | l| l| c | c c c|c}
\toprule
 & $\hat{\mathbf{L}}$ & M & H & $G_{..,x}$ & $G_{..,y}$ & $G_{..,z}$ & Note \\
\hline
\midrule
MnO &
PM  &
&
$O_h$ &
$0$
&
$0$
&
$0$\\
MnO &
[111] &
$\bar{3}m1'$
&
$D_{3d}$ &
$\left[
\begin{array}{ccc}
0 & -a & a \\
-a & b & 0 \\
a & 0 & -b \\
\end{array}
\right]$
&
$\left[
\begin{array}{ccc}
-b & a & 0 \\
a & 0 & -a \\
0 & -a & b \\
\end{array}
\right]$
&
$\left[
\begin{array}{ccc}
b & 0 & -a \\
0 & -b & a \\
-a & a & 0 \\
\end{array}
\right]$\\
MnO &
\parbox{0.2cm}{
$[1\bar{1}0]$, \\
$[11\bar{2}]$}
&
$2/m.1'$
&
$C_{2h}$ &
$\left[
\begin{array}{ccc}
d+f+g & f-g & c+h \\
f-g & f-d+g & h-c \\
c+h & h-c & e \\
\end{array}
\right]$
&
$\left[
\begin{array}{ccc}
d-f-g & -f+g & c-h \\
-f+g & -d-f-g & -c-h \\
c-h & -c-h & -e \\
\end{array}
\right]$
&
$\left[
\begin{array}{ccc}
b & 0 &  a \\
0 & -b & -a \\
a & -a & 0 \\
\end{array}
\right]$
&
$C_2\parallel [1\bar{1}0]$\\
MnO$^*$ &
$[1\bar{1}0]$ & N/A &
$D_{2h}$ &
$\left[
\begin{array}{ccc}
0 & 0 & c+h \\
0 & 0 & h-c \\
c+h & h-c & 0 \\
\end{array}
\right]$
&
$\left[
\begin{array}{ccc}
0 & 0 & c-h \\
0 & 0 & -c-h \\
c-h & -c-h & 0 \\
\end{array}
\right]$
&
$\left[
\begin{array}{ccc}
b & 0 &  0 \\
0 & -b &  0\\
0 & 0 & 0 \\
\end{array}
\right]$
&
$C_2\parallel [1\bar{1}0]$\\
\parbox{0.2cm}{
MnF$_2$,\\ NiF$_2$
}
 &
PM &

&
$D_{4h}$ &
$\left[
\begin{array}{ccc}
0 & 0 & 0 \\
0 & 0 & a \\
0 & a & 0 \\
\end{array}
\right]$
&
$\left[
\begin{array}{ccc}
0 & 0 & -a \\
0 & 0 & 0 \\
-a & 0 & 0 \\
\end{array}
\right]$
&
$0$\\
NiF$_2$ &
[010] &
$m'm'm$
&
$C_{2h}$ &
$\left[
\begin{array}{ccc}
c & 0 & 0 \\
0 & d & a \\
0 & a & e \\
\end{array}
\right]$
&
$
\left[
\begin{array}{ccc}
0 & f & -a' \\
f & 0 & 0 \\
-a' & 0 & 0 \\
\end{array}
\right]$
&
$\left[
\begin{array}{ccc}
0 & b & g \\
b & 0 & 0 \\
g & 0 & 0 \\
\end{array}
\right]$
&
$C_2\parallel x$\\
MnF$_2$ &
[001] &
$4'/mm'm$
&
$D_{2h}$ &
$\left[
\begin{array}{ccc}
0 & 0 & 0 \\
0 & 0 & a \\
0 & a & 0 \\
\end{array}
\right]$
&
$
\left[
\begin{array}{ccc}
0 & 0 & -a' \\
0 & 0 & 0 \\
-a' & 0 & 0 \\
\end{array}
\right]$
&
$\left[
\begin{array}{ccc}
0 & b & 0 \\
b & 0 & 0 \\
0 & 0 & 0 \\
\end{array}
\right]$\\
\parbox{1cm}{
MnTe, \\CrSb 
}&
PM &
&
$D_{6h}$ &
$\left[
\begin{array}{ccc}
0 & 0 & 0 \\
0 & 0 & a \\
0 & a & 0 \\
\end{array}
\right]$
&
$\left[
\begin{array}{ccc}
0 & 0 & -a \\
0 & 0 & 0 \\
-a & 0 & 0 \\
\end{array}
\right]$
&
$0$
& \footref{ftn1}\\
CrSb &
[0001] &
$6'/m'm'm$
&
$D_{3d}$ &
$\left[
\begin{array}{ccc}
-b & 0 & 0 \\
0 & b & a \\
0 & a & 0 \\
\end{array}
\right]$
&
$\left[
\begin{array}{ccc}
0 & b & -a \\
b & 0 & 0 \\
-a & 0 & 0 \\
\end{array}
\right]$
&
$0$
&
$C_2\parallel x$\footnote{\label{ftn1} X-axis points along
Cr-Sb projection in the basal plane.}\\
MnTe &
$[11\bar{2}0]$
 &
 $mmm.1$
 &
$D_{2h}$ &
$\left[
\begin{array}{ccc}
0 & 0 & 0 \\
0 & 0 & a' \\
0 & a' & 0 \\
\end{array}
\right]$
&
$
\left[
\begin{array}{ccc}
0 & 0 & -a \\
0 & 0 & 0 \\
-a & 0 & 0 \\
\end{array}
\right]$
&
$\left[
\begin{array}{ccc}
0 & -c & 0 \\
-c & 0 & 0 \\
0 & 0 & 0 \\
\end{array}
\right]$
& 
$ \mathbf{L}\perp x$\footref{ftn1}\\
MnTe &
$[1\bar{1}00]$ &
$m'm'm$
&
$C_{2h}$ &
$\left[
\begin{array}{ccc}
0 & 0 & d \\
0 & 0 & a \\
d & a & 0 \\
\end{array}
\right]$
&
$
\left[
\begin{array}{ccc}
0 & 0 & -a' \\
0 & 0 & e \\
-a' & e & 0 \\
\end{array}
\right]$
&
$\left[
\begin{array}{ccc}
f & c & 0 \\
c & g & 0 \\
0 & 0 & h \\
\end{array}
\right]$
&
\parbox{1cm}{
$C_2\parallel z$, \\ $\mathbf{L}\parallel x$}
\\
\bottomrule
\end{tabular}
\end{table*}

\begin{table*}[ht]
\centering
\caption{Transformation of $G$ under $\mathbf{L}\rightarrow - \mathbf{L}$}
\label{tab:L-inverse}
\renewcommand{\arraystretch}{1.1}
\begin{tabular}{l | l | c | c c |c}
\toprule
 & $\hat{\mathbf{L}}$ & Group &  & unitary operation  & Note \\
\hline
\midrule
MnO &
$D_{3d}$ &
id
&
id
&
id
\\
MnO &
\parbox{0.2cm}{
$(1\bar{1}0)$, \\
$(11\bar{2})$} &
$C_{2h}$ &
id
&
id
&
$C_2\parallel [1\bar{1}0]$\\
NiF$_2$ &
(010) &
$C_{2h}$ &
$\left( 
\begin{array}{rrrrrrrr}
a & a' & b & c & d & e & f & g \\
a & a' & b & -c & -d & -e & -f & -g \\
\end{array}
\right)$
&
$ \sigma_{001}$

&
$C_2\parallel x$\\
MnF$_2$ &
(001) &
$D_{2h}$ &
$\left(
\begin{array}{llr}
a & a' & b \\
a' & a & -b \\
\end{array}
\right)$
&
$ \sigma_{110}$
\\
CrSb &
(0001) &
$D_{3d}$ &
$\left(
\begin{array}{rr}
a & b \\
a & -b 
\end{array}
\right)$
&
$ \sigma_{(01\bar{1}0)}$
&
$C_2\parallel x$\footnote{\label{ftn1} X-axis points along
Cr-Sb projection in the basal plane.}\\
MnTe &
$(11\bar{2}0)$
 &
$D_{2h}$ &

&
none
& 
$ \mathbf{L}\perp x$\footref{ftn1}\\
MnTe &
$(1\bar{1}00)$ &
$C_{2h}$ &
$\left(
\begin{array}{rrrrrrrr}
a & a' & c & d & e & f & g & h \\
a & a' & c & -d & -e & -f & -g & -h \\
\end{array}
\right)$
&

$\sigma_{(01\bar{1}0)}$
&
\parbox{1cm}{
$C_2\parallel z$, \\ $\mathbf{L}\parallel x$
}
\\
\bottomrule
\end{tabular}
\end{table*}
\begin{figure}
    \centering
    \includegraphics[width=0.9\columnwidth]{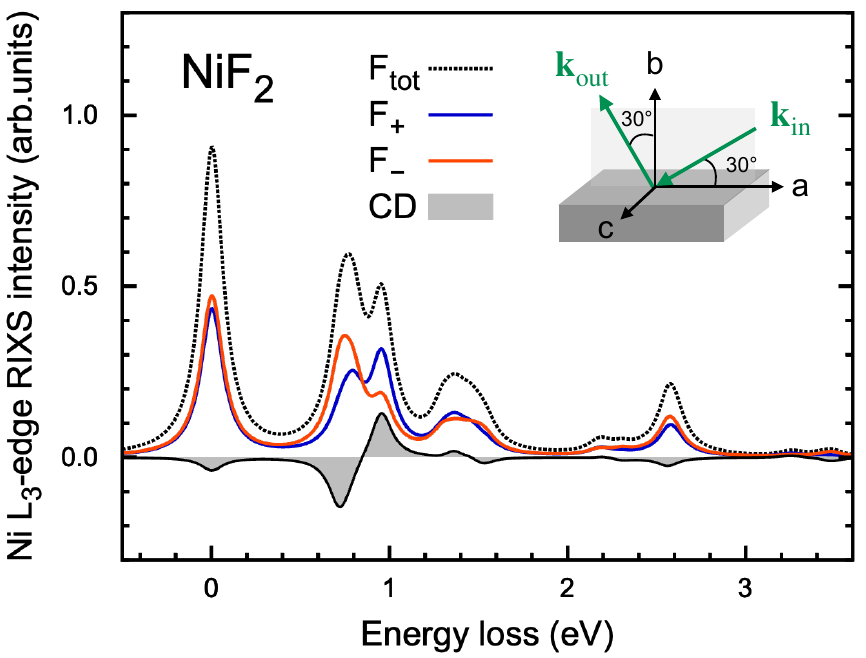}
\caption{Ni $L_3$-edge RIXS intensities computed with left- (blue) and right- (red) circularly polarized x-rays, and the CD (gray) for the altermagnetic NiF$_2$ with ${\hat{\bf L}} = [010]$. The used geometry is shown in the inset.}
    \label{fig_nif2}
\end{figure}
\begin{figure*}[t]
\includegraphics[width=2.00\columnwidth]{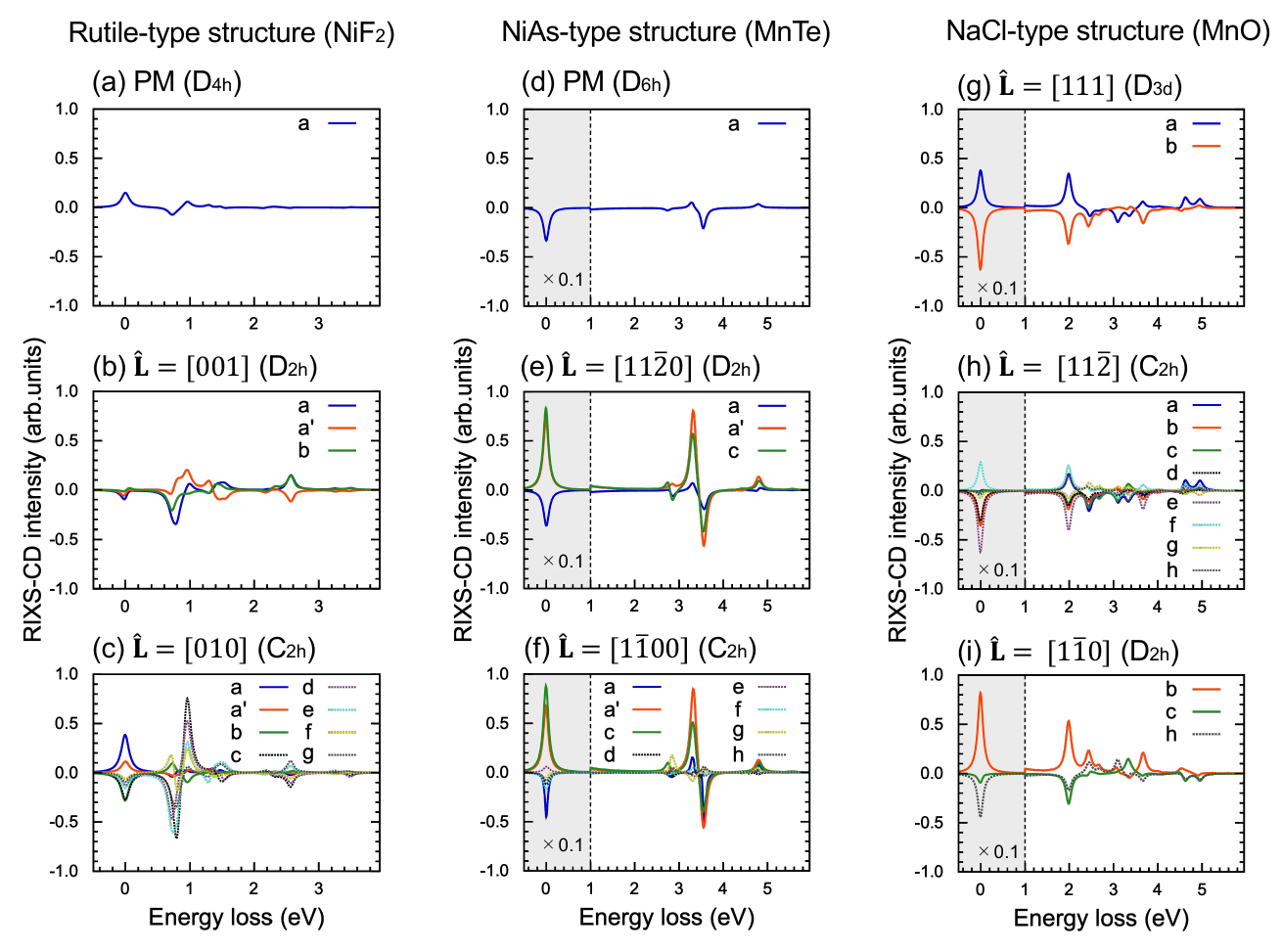}
\caption{$L_3$-edge RIXS-CD tensors calculated for selected materials (NiF$_2$, MnTe, and MnO). The incident photon energy $\omega_{\rm in}$ is set to the energy corresponding to the maximum absorption intensity at the $L_3$ edge. (a) NiF$_2$ in the PM phase ($D_{4h}$ unitary subgroup), (b) the altermagnetic phase with $\hat{\bL} = [001]$ ($D_{2h}$), and (c) the altermagnetic phase with $\hat{\bL} = [010]$ ($C_{2h}$); (d) MnTe in the PM phase ($D_{6h}$), (e) the altermagnetic phase with $\hat{\bL} = [11\overline{2}0]$ ($D_{2h}$), and (f) the altermagnetic phase with $\hat{\bL} = [1\overline{1}00]$ ($C_{2h}$); (g) MnO in the antiferromagnetic phase with $\hat{\bL} = [111]$ ($D_{3d}$), (h) the antiferromagnetic phase with $\hat{\bL} = [11\overline{2}]$ ($C_{2h}$), and (i) $\hat{\bL} = [1\overline{1}0]$ ($D_{2h}$).  In the panels, the spectral intensities in the gray-shaded regions are multiplied by the indicated factors.}
\label{fig_rixs_cd}
\end{figure*}

\section{Results}
\subsection{Symmetry analysis}
In the impurity approximation each atom contributes independently of the others and the RIXS amplitude is insensitive of the sublattice translations. Moreover,
due to the broken time-reversal symmetry in the RIXS process it is only the unitary (halving) subgroup of the magnetic point group that determines the form of the $G_{\alpha\beta,\eta}$ tensor. In Table~\ref{tab:symmetry-data} we list the magnetic point groups $\bM$ and the unitary subgroups $\mathbf{H}$ for several materials of interest~\footnote{In the paramagnetic state the unitary subgroup is the crystallographic point group and the magnetic point group, not shown in the table, is the corresponding grey group.}.
To determine the form of the $G_{\alpha\beta,\gamma}$ tensor we look for the invariants within the $\operatorname{Sym}^2 V\otimes \Lambda^2V$ representation of $\mathbf{H}$. Here $V$ is the vector representation, $\operatorname{Sym}^2V$ and
$\Lambda^2V$ are the symmetric and antisymmetric product (pseudovector) representations, respectively. The results of our analysis are shown in Table~\ref{tab:symmetry-data}.

While for the $O_h$ group the RIXS-CD is forbidden, in the lower-symmetry rutile and NiAs structures the RIXS-CD is only forbidden for specific geometries, e.g., for $\hat{\bk}_\text{in}$ along the $z$-axis, even in the paramagnetic (PM) state. In the high-symmetry structures magnetic ordering lowers the symmetry, which depends on both the 
periodicity of the ordered phase and the orientation of the local moments. This gives rise to additional non-zero elements of the $G_{\alpha\beta,\gamma}$ tensor and thus distinguishes the ordered and PM phase. 

How does the RIXS-CD behave under the reversal of the N\'eel vector $\bL\rightarrow -\bL$? To answer this question we observe that RIXS-CD is invariant under translation as well as under inversion. In classical antiferromagnets, the $\bL$-reversal amounts to applying either of the operations and thus the $G_{\alpha\beta,\gamma}$ tensor remains unchanged.
This is the case of MnO in Table~\ref{tab:L-inverse}. In altermagnets, there are two possibilities: i) the states with $\bL$ and $-\bL$ are connected by a unitary 
operation, i.e., an operation from the crystallographic space group of the PM state, which does not belong to the magnetic space group of the ordered state or ii)
there is no unitary map between $\bL$ and $-\bL$. Since the unitary subgroup $H$ is the same for $\bL$ and $-\bL$,
the $G_{\alpha\beta,\gamma}$ tensor for both states has the form given by Table~\ref{tab:symmetry-data}. 
In case (i) there is a mapping between the matrix elements of $G_{\alpha\beta,\gamma}(\bL)$ and $G_{\alpha\beta,\gamma}(-\bL)$
as shown in Table~\ref{tab:L-inverse}. 
In case (ii) the matrix elements for $\bL$ and $-\bL$ are independent from each other. An example of the latter is MnTe with $\bL$ along the $[11\bar{2}0]$ direction. We put this somewhat counterintuitive results to a numerical test in the following section.
We point out that similar consideration applies also to ferromagnets. 

The above analysis assumes a fixed flux of incoming photons. However, the circular dichroism in experiments is usually calibrated by the total scattering/absorption intensity. In Table~\ref{tab:tot} we provide the form of the $G'_{\alpha\beta,\gamma\delta}$ tensor for the high symmetry structures.

\begin{figure*}
    \centering
    \includegraphics[width=1.8\columnwidth]{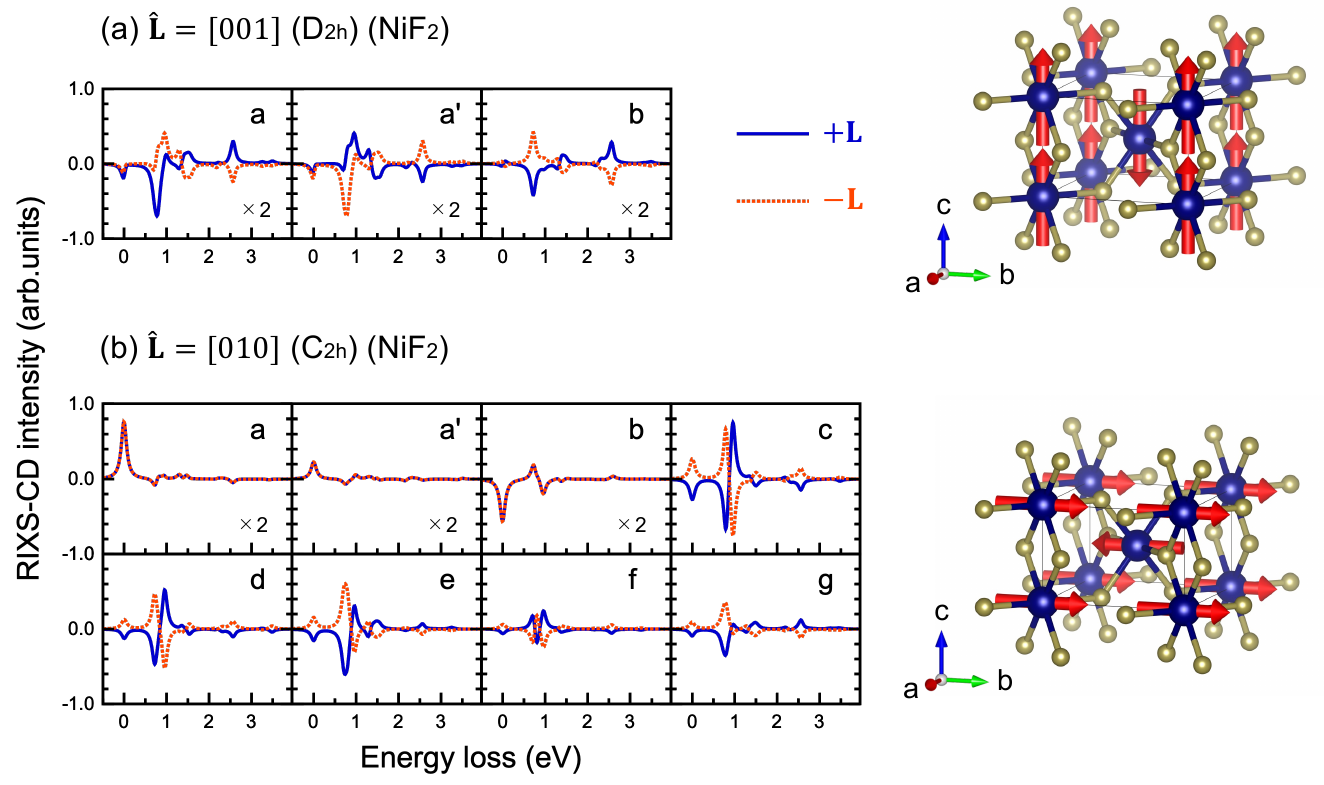}
\caption{Ni $L_3$-edge RIXS-CD tensors computed for $+\bL$ (red) and $-\bL$ (blue) in the altermagnetic phase of NiF$_2$ with (a) $\hat{\bL} = [001]$ and (b) $\hat{\bL} = [010]$. Scaling by a factor of 2 is applied to the panels showing tensor elements $a$, $a'$, $b$.}
    \label{fig_neelvec_rutile}
\end{figure*}

\begin{figure*}
    \centering
    \includegraphics[width=1.8\columnwidth]{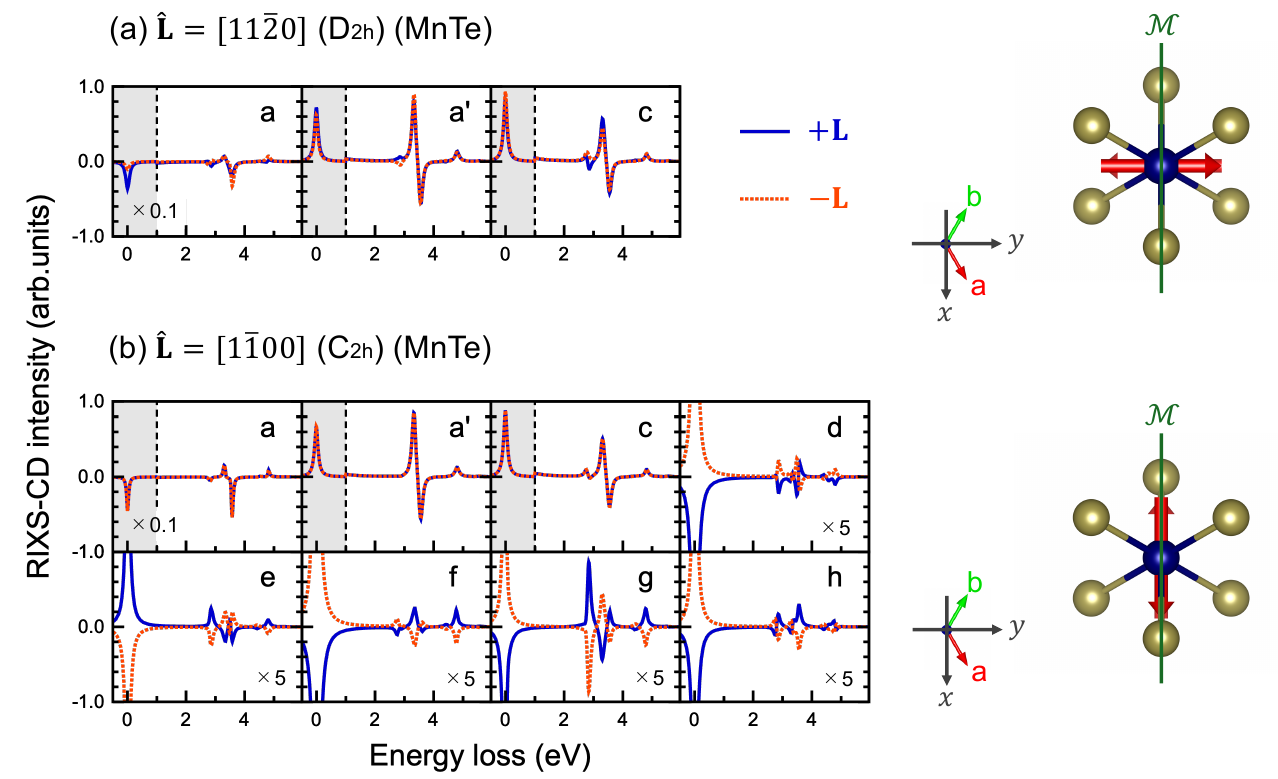}
\caption{Mn $L_3$-edge RIXS-CD tensors computed for $+\bL$ (red) and $-\bL$ (blue) in the altermagnetic phase of MnTe with (a) $\hat{\bL} = [11\overline{2}0]$ and (b) $\hat{\bL} = [1\overline{1}00]$. In the panels, the spectral intensities in the gray-shaded elastic-peak regions are multiplied by a factor of 0.1. A factor of 5 is applied to the panels showing tensor elements $d$–$h$. The N\'eel states corresponding to cases (a) and (b) are also shown.}
    \label{fig_neelvec_nias}
\end{figure*}

\begin{figure}
    \centering
    \includegraphics[width=0.9\columnwidth]{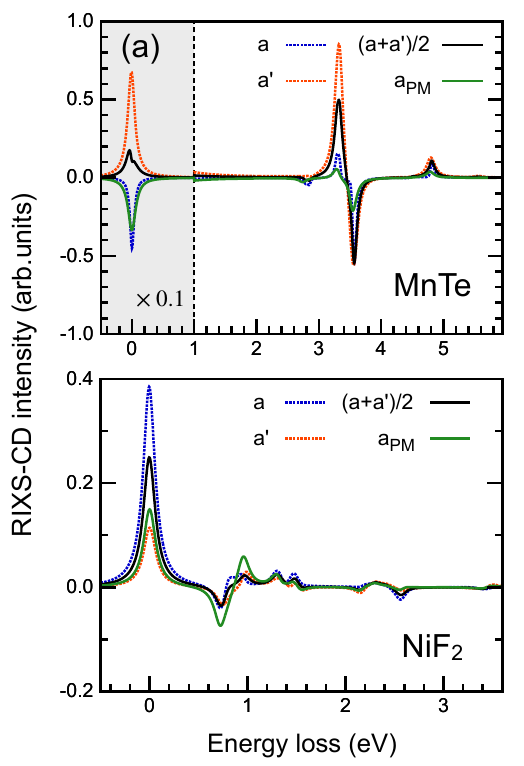}
\caption{Tensor components $a$ (blue) and $a'$ (red) in the altermagnetic phase, and $a$ (green) in the paramagnetic (PM) phase for (a) MnTe with $\bL = \langle1\overline{1}00\rangle$ order and (b) NiF$_2$ with $\bL = \langle 010\rangle$ order. The average of the $a$ and $a'$ components in the altermagnetic phase is shown as a black line.}
    \label{fig_rutile_afl001_pm}
\end{figure}

\subsection{Selected materials}

In the following, we evaluate the RIXS-CD tensor for selected materials:~NiF$_2$, MnTe, and MnO.
The corresponding crystal structures are shown in Fig.~\ref{fig_geo}(b–d).
The RIXS intensities are simulated using an atomic model that represents the x-ray-excited metal site, constructed following Refs.~\onlinecite{Hariki24,Hariki25}. Additional computational details are provided in Appendix~\ref{method}. To simulate the magnetically ordered phase, a small molecular field is applied to align the N\'eel vector $\bL$ along a specified direction, and the RIXS signals from individual sites in the magnetic unit cell are summed~\cite{Winder20,Hariki24}. By construction, the atomic models fix the $d$-electron count to the formal valency $d^n$ ($d^8$ for NiF$_2$ and $d^5$ for MnTe and MnO), thereby restricting the configuration evolution to the $d^n \rightarrow \underline{c}d^{n+1} \rightarrow d^n$ in the $L$-edge (2$p$--3$d$) RIXS process, where charge transfer with surrounding ions is prohibited. Here, $\underline{c}$ denotes a 2$p$ core hole. Thus the models capture only excitations bounded at the x-ray excited site. This approach is routinely used to simulate $dd$ excitations in the RIXS spectra, including crystal-field and Coulomb multiplet excitations, in correlated insulators~\cite{Ament11,Mitrano24,deGroot24,groot_kotani,Tomiyasu17,Hunault18}. 
%
%
Numerical studies on MnTe revealed a semiquantitative agreement between the XMCD spectra obtained with atomic model and Anderson-impurity-model DFT+DMFT treatment~\cite{Hariki24}. A typical set-up and results for total RIXS and RIXS-CD is shown in Fig.~\ref{fig_nif2}, with NiF$_2$ as an example.

\subsubsection{MnO}
\label{subsub:mno}

MnO crystallizes in the NaCl structure and orders antiferromagnetically in the G-type structure with parallel
arrangement of moments within (111) planes~\cite{Roth1958}. More recent neutron study~\cite{Goodwin2006} identified
the $\langle 11\bar{2}\rangle$ as the easy axis. Here, we consider three high symmetry directions of the
N\'eel vector $\bL$ $[111]$, $[1\bar{1}0]$ and $[11\bar{2}]$. 
The corresponding RIXS-CD tensors are shown in Fig.~\ref{fig_rixs_cd}(g–i).
The PM phase with $O_h$ does not admit 
RIXS-CD. The magnetic ordering reduces the symmetry in two ways: i) by distributing originally equivalent atoms on
magnetic sublattices regardless of the moment direction, described by the spin group. ii) by selecting specific
direction of the local moments, described by the magnetic group. In general, the spin group
constrains the possible magnetic groups, however, this is not necessarily the case in the atomic approximation
as we explain below. 

For $\bL\parallel [111]$ the orientation of the magnetic moments does not further reduce the unitary subgroup $H=D_{3d}$
imposed by the G-type order. For $\bL\parallel [11\bar{2}]$ the unitary subgroup $H=C_{2h}$. In these two cases,
adding a magnetic moment with the desired orientation to an atom in the cubic crystal field, as done in the atomic
approximation, results in the same group $H$.

For $\bL\parallel [1\bar{1}0]$ this is not the case. Taking into account both the lattice symmetry (i) and magnetic moment
orientation (ii) we end up with $H=C_{2h}$. However, if we only add the local moments to atoms in cubic crystal field
we get $H=D_{2h}$, which is obviously incompatible with the G-type order on the lattice. Is this a useful approximation?
How is this mismatch possible? This mismatch comes from the fact that we kept the crystal field cubic even though
the sounding lattice adopted $D_{3d}$ symmetry. In the impurity approximation there are two ways how the symmetry is lowered
due to the magnetic ordering. The atoms shift in response to the magnetic order and this is reflected in the crystal field.
The coupling between the impurity and the rest of the lattice~\footnote{Hybridization function in the Anderson impurity model.},
neglected in the atomic approximation, reflects the lattice $D_{3d}$ symmetry. If these effects can be viewed as small perturbations
then the decomposition into the leading terms with $D_{2h}$ symmetry, shown in Table~\ref{tab:symmetry-data}, and corrections with $C_{2h}$ symmetry may beneficial. In Appendix B we introduce a small trigonal distortion to the Mn crystal field
and show the corresponding irreducible spectra with the expected symmetry.


\subsubsection{MnF$_2$, NiF$_2$}

These two materials belong to a well studied antiferromagnetic insulators with rutile structure, i.e., altermagnets.
The RIXS-CD is allowed even in the PM state with $H=D_{4h}$, but is forbidden for light coming along the 
$c$-axis. The two materials are distinguished by the easy-axis orientation. 

Below the transition temperature the moments
in MnF$_2$  align along the $c$-axis~\cite{Erickson1953}. The XMCD is not allowed~\cite{Hariki24_mnf2,Hariki24_ruo2}, while RIXS-CD is allowed for any incoming light direction. For the light coming along
a crystallographic axis there are two nodal planes perpendicular to the remaining axes where RIXS-CD vanishes.
There are two possible domains $[001]$ and $[00\bar{1}]$, which are related by a unitary operation, see Table~\ref{tab:L-inverse}. 
We have not performed numerical simulations from MnF$_2$. Instead, we have calculated the spectra for NiF$_2$ in 
a hypothetical $\langle 001\rangle$ structure in order to numerically test our symmetry analysis, Fig.~\ref{fig_rixs_cd}b.


NiF$_2$ is the only member of the $3d$ difluoride series with the easy axis in the $ab$-plane, $\bL=\langle 100\rangle$~\cite{Erickson1953,Matarrese54}.
The XMCD is allowed with the Hall vector in the $ab$-plane and perpendicular to $\bL$~\cite{Hariki25}. The moment reversal is facilitated
by $(001)$ mirror plane, see Table~\ref{tab:L-inverse}. The spectra for the irreducible contributions to the $G$-tensor  
are shown in Fig.~\ref{fig_rixs_cd}(a-c) and the effect on $\bL$- reversal
in  NiF$_2$ is shown in Fig.~\ref{fig_neelvec_rutile}. 

\subsubsection{MnTe,CrSb}

MnTe and CrSb are much studied altermagnets with NiAs structure~\cite{Smejkal22}. The PM $H=D_{6h}$ leads to the same form of G-tensor as in the rutile structure with a single irreducible spectral function and no RIXS-CD for the light coming along the $c$-axis. 
The RIXS-CD tensors calculated for MnTe are given in Fig.~\ref{fig_rixs_cd}(d–f).

The easy axis of CrSb~\cite{Park20} is along the $c$-axis and leads to $H=D_{3d}$. The XMCD is forbidden in this symmetry as is the
RIXS-CD for the light coming along the $c$-axis. Compared to the PM state there is a second irreducible spectral function. The states with $\bL$ and $-\bL$ are related by a vertical mirror plane (Table~\ref{tab:L-inverse}). We did not calculate the irreducible
spectral functions since metallic material requires full AIM treatment rather than the atomic model used in this work.

The easy axis of MnTe is in the $\langle 1\bar{1}00\rangle$ direction~\cite{Kriegner17}. This allows XMCD with the Hall vector along the
$c$-axis~\cite{Gonzalez23,Hariki24}. The $H=C_{2h}$ symmetry leads to 8 irreducible spectral functions, see Table~\ref{tab:symmetry-data} and Fig.~\ref{fig_rixs_cd}(f). The states with $\bL$ and $-\bL$ are related by a vertical mirror plane, which contains $\bL$,
as summarized in Table~\ref{tab:L-inverse} and numerically demonstrated in Fig.~\ref{fig_neelvec_nias}(b). 

In MnTe it is relatively easy to rotate the N\'eel vector within the $ab$-plane~\cite{Kriegner17}.
The $\bL=\langle 11\bar{2}0\rangle$ is another high symmetry direction, which yields $H=D_{2h}$. The XMCD is forbidden~\cite{Hariki24}, but there are 3 RIXS-CD irreducible spectral functions. Interestingly, 
the states with $\bL$ and $-\bL$ are not
related by any unitary transformation and thus the corresponding RIXS-CD spectra are not related to one another as is demonstrated by numerical simulation in Fig.~\ref{fig_neelvec_nias}(a). 


\subsection{Domain structure with RIXS-CD}
Spectroscopic investigations of ordered magnets may be plagued by the existence of multiple domains within the beam spot~\cite{Amin24,Hariki24}.
Averaging over an even population of domains recovers the PM symmetry. It is difficult to distinguish a domains-averaged  magnet
from a paramagnet in x-ray absorption since the XMCD vanishes and the total absorption is usually rather insensitive to the magnetic order.
Is it possible to do so with RIXS-CD? We consider two cases: (i) MnTe with experimental $\langle 1\bar{1}00\rangle$ easy axis with six domains
and (ii) 
NiF$_2$ with $\bL=\langle 010 \rangle$ order. 
In MnTe an equal population of time-reversed domains with $\bL$ and $-\bL$ leads to vanishing of all irreducible functions, but $a$, $a'$ and $c$.
Further averaging over the 120$^\circ$ domains eliminates $c$ and averages over $a$ and $a'$. The comparison of domain averaged and PM spectra 
are shown in Fig.~\ref{fig_rutile_afl001_pm}a. While the shape of the spectra is similar, the CD magnitude several times larger in the
magnetically ordered state than in the PM state. A qualitative difference--opposite sign--is found for the elastic peak, however, we do not expect
the impurity approximation to perform well in this region. In Fig.~\ref{fig_rutile_afl001_pm}b we show the comparison for NiF$_2$.
The averaging over time-reversed domains with $\bL$ and $-\bL$ leads to vanishing of all irreducible functions, but $a$, $a'$ and $b$. While the $90\deg$ rotation eliminates $b$ leaving a single irreducible spectral function 
 $a_{\text{avg}}(\omega)=(a(\omega)+a'(\omega))/2$. 
Again we find similar shapes of $a_{\text{avg}}(\omega)$ and $a_{\text{PM}}(\omega)$, but now the magnitude, with the exception of the elastic peak, is larger 
in the PM phase. These results suggest that there is a practically observable difference between the PM and
magnetically ordered state even if domains are present, however, the nature of this difference is material specific.
If the present theory describes the real material with sufficient accuracy it shall be
even possible to obtain the domain weights by fitting the $\bk_\text{out}$ and $\bk_\text{in}$ dependency of RIXS-CD.

\section{Conclusions}
We have provided symmetry analysis and numerical simulations of RIXS-CD in several common structures. We have employed the impurity approximation
which can be viewed as a high-symmetry limiting case in the sense that if RIXS-CD is allowed for a given geometry in the impurity approximation
it is also allowed in exact treatment, which involves lower symmetry states with finite momenta. Since the Kramers–Heisenberg formula is not invariant under
anti-unitary symmetries of the Hamiltonian, the RIXS spectra do not reflect the time-reversal symmetry breaking due to magnetic ordering. 
Only possible breaking of unitary symmetries in the ordered phase is reflected in the RIXS spectra. As a result RIXS-CD may be finite even
in the normal phase. Another interesting consequence is the relationship between spectra from states with reversed magnetic moments. In classical 
antiferromagnets the time-reversal does not change the RIXS-CD spectra since this is equivalent to translation or inversion operation.
In altermangets, the RIXS-CD spectra for $\bL$ and $-\bL$ are independent from one another in general unless there is a unitary operation 
mapping $\bL$ to $-\bL$ in the symmetry group of the normal state. The latter can happen for $\bL$ pointing along the high-symmetry direction 
in the structure.

The RIXS-CD dependence to the wave vectors of the incoming and outgoing photons originates from constrains on photon polarization and
from the momentum-conservation constrain on the final state. In the impurity approximation only the former is taken into account. This may be a reasonable approximation for heavy, e.g., crystal-field excitations, but ultimately has to be verified by comparison to experiment.
While no propagating excitations such as magnons are allowed in the impurity approximation we find that RIXS-CD is rather prolific. This implies
that interpretation of RIXS-CD for example in terms of chirality of excitations such as magnons must be approached with caution.

\begin{acknowledgements}
We thank Jakub \v{Z}elezn\'y, Hakuto Suzuki, and Anna Kauch for discussions and critical reading of the manuscript. This work was supported by JSPS KAKENHI Grant Numbers 25K00961, 25K07211, 23H03816, 23H03817, the 2025 Osaka Metropolitan University (OMU) Strategic Research Promotion Project (Young Researcher) (A.H.),
by the project Quantum materials for applications in sustainable technologies (QM4ST),
funded as project No. CZ.02.01.01/00/22 008/0004572 by Programme Johannes Amos Commenius,
call Excellent Research, and by the Ministry of Education, Youth and Sports of the Czech Republic 
through the e-INFRA CZ (ID:90254).
\end{acknowledgements}

\begin{figure}[t]
    \centering
    \includegraphics[width=0.9\columnwidth]{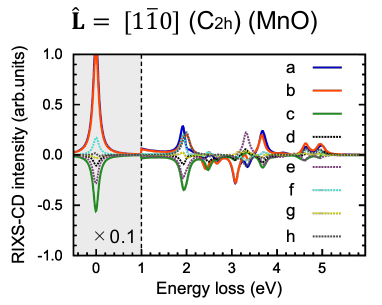}
\caption{Irreducible RIXS-CD densities for MnO with $\hat{\bL} = [1\overline{1}0]$, including a small trigonal distortion that introduces a 0.2~eV splitting between $e_{g\pi}$ and $a_{1g}$ levels in the $t_{2g}$ manifold.}
    \label{fig_nacl_l1-10_c2h}
\end{figure}

\appendix

\section{RIXS-CD simulation}
\label{method}

We simulate RIXS intensities using an atomic model constructed following Refs.~\onlinecite{Hariki24,Hariki25}. The atomic Hamiltonian includes the valence–valence interaction within the 3$d$ shell, the core–valence interaction between the 3$d$ and 2$p$ shells in the intermediate state of the RIXS process, spin–orbit coupling in both the 3$d$ and 2$p$ shells, and the crystal-field splitting of the 3$d$ levels. The Hamiltonian is diagonalized numerically to obtain the full eigenvalue spectrum of the initial, intermediate, and final states, followed by evaluation of the RIXS intensities using the Kramers–Heisenberg formula. The inverse lifetime constant $\Gamma$ is set to 0.3~eV throughout the present study. The crystal-field parameters are derived from DFT calculations using the Wien2k and Wannier90 packages~\cite{wien2k,wannier90}, performed for the experimental crystal structures of the simulated compounds. The Slater integrals for the 2$p$--3$d$ interaction are obtained from atomic Hartree–Fock calculations and reduced to 80~\% of their atomic values to account for the effects of higher configurations neglected in the atomic treatment, a well-established approach for simulating core-level spectra at 3$d$ transition metal edges~\cite{Hariki2017,Matsubara05,Sugar72,Tanaka92,Groot90}. The valence–valence (3$d$--3$d$) interaction is parameterized by the Hund's coupling $J = (F_2 + F_4)/14$. The parameters used for MnTe and NiF$_2$ are provided in Refs.~\onlinecite{Hariki24} and~\onlinecite{Hariki25}, respectively. For MnO, we employ $J = 0.81$~eV, a typical value for Mn oxides.

\section{MnO $\bL=[1\bar{1}0]$ in $C_{2h}$}
The magnetic ordering with $[111]$ propagation vector enforced $D_{3d}$ or lower symmetry (depending on $\bL$). The reduction of the PM cubic symmetry
is reflected in the electronic band structure as well as lattice deformation, which feeds back to the electronic structure. To mimic the latter effect
we have added a trigonal crystal field of 0.2~eV and repeated the RIXS-CD calculation for $\bL=[1\bar{1}0]$. The corresponding irreducible
spectral densities, which follow the expectations for the $C_{2h}$ symmetry are shown in Fig.~\ref{fig_nacl_l1-10_c2h}.

\section{Total RIXS intensity}
In Table~\ref{tab:tot} we show the irreducible components of the total (unpolarized in, unpolarized out) RIXS for the discussed
systems. We do not show the components for the magnetically ordered phases of MnO where we find it of little practically due
to the large number of distinct spectral functions.

\begin{table}[ht]
\centering
\caption{The $G'$ tensor in  Voigt notation $G'_{\alpha\beta,\gamma\delta}=G^\text{Voigt}_{i(\alpha\beta)
i(\gamma\delta)}$ with the labeling $i(xx)=1, i(yy)=2, i(zz)=3, i(yz)=i(zy)=4, i(xz)=i(zx)=5, i(xy)=i(yz)=6.$}
\label{tab:tot}
\renewcommand{\arraystretch}{1.1}
\begin{tabular}{l | c | c | l |c}
\toprule
 & $\hat{\mathbf{L}}$ & H & $G^\text{Voigt}$ & Note \\
\hline
\midrule
MnO &
PM  &
$O_h$ &
$\left[
\begin{array}{cccccc}
a & b & b &0 & 0 & 0\\
b& a & b & 0 & 0 & 0 \\
b & b & a & 0 & 0 & 0\\
0 & 0 & 0 & c & 0 & 0\\
0 & 0 & 0 & 0 & c & 0 \\
0 & 0 & 0 & 0 & 0 & c\\
\end{array}
\right]$
\\
\parbox{0.2cm}{
MnF$_2$,\\ NiF$_2$
}
 &
PM &
$D_{4h}$ &
$\left[
\begin{array}{cccccc}
a & e & c &0 & 0 & 0\\
e & a & c & 0 & 0 & 0 \\
d & d & b & 0 & 0 & 0\\
0 & 0 & 0 & f & 0 & 0\\
0 & 0 & 0 & 0 & f & 0 \\
0 & 0 & 0 & 0 & 0 & g\\
\end{array}
\right]$
&

\\
NiF$_2$ &
(010) &
$C_{2h}$ &
$\left[
\begin{array}{cccccc}
. & . & . & . & 0 & 0\\
. & . & . & . & 0 & 0 \\
. & . & . & . & 0 & 0\\
. & . & . & . & 0 & 0\\
0 & 0 & 0 & 0 & . & . \\
0 & 0 & 0 & 0 & . & .\\
\end{array}
\right]$
&

$C_2\parallel x$\\
MnF$_2$ &
(001) &
$D_{2h}$ &
$\left[
\begin{array}{cccccc}
a & d & e &0 & 0 & 0\\
g & b & f & 0 & 0 & 0 \\
h & p & c & 0 & 0 & 0\\
0 & 0 & 0 & q & 0 & 0\\
0 & 0 & 0 & 0 & r & 0 \\
0 & 0 & 0 & 0 & 0 & s\\
\end{array}
\right]$
&

\\
\parbox{1cm}{
MnTe, \\CrSb 
}&
PM &
$D_{6h}$ &
$\left[
\begin{array}{cccccc}
a+f & a-f & c &0 & 0 & 0\\
a-f & a+f & c & 0 & 0 & 0 \\
d & d & b & 0 & 0 & 0\\
0 & 0 & 0 & e & 0 & 0\\
0 & 0 & 0 & 0 & e & 0 \\
0 & a & 0 & 0 & 0 & f\\
\end{array}
\right]$
&

\\
CrSb &
(0001) &
$D_{3d}$ &
$\left[
\begin{array}{cccccc}
a+f & a-f & c &0 & h & 0\\
a-f & a+f & c & 0 & -h & 0 \\
d & d & b & 0 & 0 & 0\\
0 & 0 & 0 & e & 0 & g\\
g & -g & 0 & 0 & e & 0 \\
0 & a & 0 & h & 0 & f\\
\end{array}
\right]$
&
$C_2\parallel x$\footnote{\label{ftn1} X-axis points along
Cr-Sb projection in the basal plane.}\\
MnTe &
$(11\bar{2}0)$
 &
$D_{2h}$ &
$\left[
\begin{array}{cccccc}
a & d & e &0 & 0 & 0\\
g & b & f & 0 & 0 & 0 \\
h & p & c & 0 & 0 & 0\\
0 & 0 & 0 & q & 0 & 0\\
0 & 0 & 0 & 0 & r & 0 \\
0 & 0 & 0 & 0 & 0 & s\\
\end{array}
\right]
$
&

$ \mathbf{L}\parallel x$\footref{ftn1}\\
MnTe &
$(1\bar{1}00)$ &
$C_{2h}$ &
$\left[
\begin{array}{cccccc}
. & . & . & 0 & 0 & .\\
. & . & . & 0 & 0 & . \\
. & . & . & 0 & 0 & .\\
0 & 0 & 0 & . & . & 0\\
0 & 0 & 0 & . & . & 0\\
. & . & . & 0 & 0 & .\\
\end{array}
\right]$
&

$C_2\parallel z$\\
\bottomrule
\end{tabular}
\end{table}

\bibliography{main}
\end{document}